\documentclass[aps, prd, twocolumn, superscriptaddress, nofootinbib]{revtex4-1}
\usepackage{graphicx}
\usepackage{dcolumn}
\usepackage{amssymb}
\usepackage{amsmath,amssymb,amsfonts}

\begin{document}

\title{Planck-scale dual-curvature lensing and spacetime noncommutativity}

\newcommand{\addressStefano}{Max-Planck-Institut fur Gravitationsphysik - Albert-Einstein-Institut
Am M\"uhlenberg 1, 14476 Potsdam-Golm, Germany}
\newcommand{\addressRoma}{Dipartimento di Fisica, Universit\`a ``La Sapienza''
and Sez. Roma1 INFN, P.le A. Moro 2, 00185 Roma, Italia}

\author{Giovanni Amelino-Camelia}
\affiliation{\addressRoma}
\author{Leonardo Barcaroli}
\affiliation{\addressRoma}
\author{Stefano Bianco}
\affiliation{\addressStefano}
\author{Laura Pensato}
\affiliation{\addressRoma}

\begin{abstract}
It was recently realized that Planck-scale momentum-space curvature, which is expected in some approaches to the
quantum-gravity problem,
can produce  dual-curvature lensing, a feature which mainly affects the direction of observation
of particles emitted by very distant sources.
Several gray areas remain in our understanding of dual-curvature lensing,
including the possibility that it might be just a coordinate artifact and the possibility
that it might be in some sense a by product of the better studied dual-curvature redshift.
We stress that data reported by the IceCube neutrino telescope should motivate a more vigorous
effort of investigation of dual-curvature lensing, and
we observe that studies of the recently proposed ``$\rho$-Minkowski noncommutative spacetime"
could be valuable from this perspective. Through a dedicated $\rho$-Minkowski analysis, we show that dual-curvature lensing is not merely
a coordinate artifact and that it can be present even in theories without dual-curvature redshift.
 \end{abstract}


\maketitle

\section{Introduction}
There has been considerable interest
(see, {\it e.g.}, Refs.\cite{grbgac,gampul,urrutia,jacobpiran,gacsmolin,gacLRR,mattiLRR})
over the last decade for the possibility that Planck-scale structures might have
observably-large implications for particle propagation over cosmological distances.
We shall here focus on one of the scenarios for the quantum-gravity realm which can motivate
such studies, the scenario such that the characteristic scale of quantum gravity (usually assumed to be of the order
of the Planck scale) plays the role of scale of curvature
of momentum space \cite{principle,corteslocrel,mignemilocrel,stringylocrel}.
The most studied effect of this sort is ``dual-curvature redshift" \cite{dualredshift},
an effect such that two ultrarelativistic
(massless or effectively massless) particles
of different energies emitted simultaneously at a distant source reach a detector at different times
(earlier discussions of this effect, before the curved-momentum-space perspective emerged, can be found,
{\it e.g.}, in Refs.\cite{grbgac,gampul,urrutia,jacobpiran,gacsmolin}). Recently it became clear that Planck-scale curvature of momentum space can also
produce the effect of ``dual-curvature lensing"
\cite{tranverseFS,tranverseNICCOLEO,tranverse3DQG}
(or ``dual-gravity lensing"\footnote{The effect is better known
as dual-gravity lensing, but actually it is present also in absence of momentum-space geometrodynamics: momentum-space
curvature is sufficient for producing the effect.
We also note that all models with Planck-scale-curved momentum space
developed so far indeed do not have momentum-space geometrodynamics, but rather have a fixed background curved momentum-space
geometry. We reserve the characterization of effects as ``gravitational" only when geometrodynamics is present.}),
an effect such that two ultrarelativistic particles
of different energies emitted by the same distant source reach a detector from different directions.

The understanding of dual-curvature lensing has not yet reached the level of our understanding
of dual-curvature redshift. Most notably previous studies establish that dual-curvature redshift is a truly physical
(observable) effect, but leave open the possibility that dual-curvature lensing might be just
a coordinate artifact. Moreover, these previous works do not clarify whether
 dual-curvature lensing is just some sort of by product
of dual-curvature redshift or instead one could have theories without dual-curvature redshift
 but with dual-curvature lensing.

 This state of affairs became recently
 more disappointing since data reported by the IceCube telescope \cite{IceCube}
 open a potentially powerful observational window
 on dual-curvature lensing, and actually a first exploratory analysis
 gave preliminarily encouraging results \cite{ryan}.

We here focus on the recently-proposed ``$\rho$-Minkowski noncommutative spacetime" \cite{tranverseNICCOLEO},
using it as a conceptual laboratory
for dual-curvature lensing. We show that $\rho$-Minkowski produces dual-curvature lensing, but dual-curvature redshift is absent.
Most importantly we establish
that in $\rho$-Minkowski
dual-curvature lensing is a truly physical effect, not merely a coordinate artifact.

\section{$\rho$-Minkowski and its relativistic properties}
We start by characterizing the relativistic properties of the
$\rho$-Minkowski noncommutative spacetime. $\rho$-Minkowski is a 3D noncommutative spacetime, characterized
by the following commutation relations among spacetime coordinates:
\begin{equation}
\label{commu-rel}
[x_i,x_0]=\rho\epsilon_{ij}x_j \ ,  \quad [x_i,x_j]=0 \ ,
\end{equation}
where $i \in\{1,2\}$.

As usual for this sort of noncommutative spacetime, in order to have an explicit description of
the relativistic properties one must adopt an ordering convention for the basis
of exponentials. We find convenient to
adopt time-to-the-right ordering, so that elements of the basis of exponentials are written as follows
$$e^{ip_jx_j}e^{ip_0x_0}$$
Relying again on experience gained working with other similar noncommutative spacetimes,
we infer  the form of the law of composition of momenta by studying the properties of products
of elements of the basis of exponentials. We have that
\begin{eqnarray}
&& e^{ip_jx_j}e^{ip_0x_0}e^{iq_lx_l}e^{iq_0x_0}\approx \label{gac1}\\
&&\,\,\,\,\,\,\,\, (1+ip_jx_j)(1+ip_0x_0)(1+iq_lx_l)(1+iq_0x_0) \approx \nonumber\\
&&\,\,\,\,\,\,\,\, (1+i(p_j+q_j-\rho p_0\epsilon_{jl}q_l)x_j)(1+i(p_0+q_0)x_0) \, ,
\nonumber
\end{eqnarray}
where we used (\ref{commu-rel}) and we focused on the leading correction in $\rho$.

The result (\ref{gac1}) suggests that, at leading order in $\rho$,
the law of composition of momenta in $\rho$-Minkowski should take the form:
\begin{equation}
\label{compo-law}
\begin{split}
(p\oplus q)_0&=p_0+q_0 \ , \\
(p\oplus q)_i &= p_i+q_i-\rho p_0\epsilon_{ij}q_j \ .
\end{split}
\end{equation}

We shall here establish that $\rho$-Minkowski allows the formulation of
a DSR-relativistic theory, in the sense first introduced in Refs.\cite{dsr1,dsr2}
(also see Refs.\cite{jurekDSRnew,leedsrPRD}),
{\it i.e.} a relativistic theory with two non-trivial relativistic invariant, also a momentum scale, in this case $\rho^{-1}$,
in addition to the speed-of-light scale (which we set to unity by choice of dimensions, but can be easily brought back into the
picture by dimensional analysis). The label ``DSR" refers to the fact that this class of relativistic
theories was at first called ``doubly-special reativity".

The DSR-compatibility of the setup requires that the action of boosts on momenta be introduced
by requiring that the deformed momentum-conservation law  be covariant.
In order to ensure this it suffices to require that a vanishing composed momentum for observer Alice, $p\oplus q=0$,
 also vanishes for observer Bob, boosted with respect to Alice:
 \begin{equation}
\label{compatibility}
\{N_i^{[p\oplus q]}, p\oplus q\}|_{p\oplus q=0}=0 \ .
\end{equation}
Taking as guidance the description of boosts in the rather similar $\kappa$-Minkowski noncommutative spacetime \cite{majRue,lukRueg,jurekDSMOMENTUM},
in the case of $\rho$-Minkowski
we are led to the following description of the action of boosts on composed momenta, when given in terms
of the action of boosts on single-particle momenta:
\begin{equation}
\label{compo-boosts}
N_i^{[p\oplus q]}=N_i^{[p]}+N_i^{[q]}-\rho p_0\epsilon_{ij}N^{[q]}_j
\end{equation}
For the action of boosts on single-particle momenta
we introduce a 6-parameter {\it ansatz}:
\begin{eqnarray}
\label{ansatz}
\{N_{i},p_{0}\} & = & p_{i} +\rho Ap_0p_i \, ,\\
\{N_{i},p_{j}\} & = & \delta_{ij}\, p_{0}+\rho\left(Bp_0^2+C\,|\vec{p}|^{2}\right)\delta_{ij}+ \nonumber \\
& & \,\,\,\,\,\,\,\, + \rho\left(Dp_0^2+E\,|\vec{p}|^{2}\right)\epsilon_{ij} +\rho F\epsilon_{il}p_lp_j \, ,
\nonumber
\end{eqnarray}
The parameters appearing in (\ref{ansatz}) can be of course determined by making use
of (\ref{compo-boosts}) and  (\ref{ansatz}) in the compatibility condition (\ref{compatibility}), finding that
the compatibility condition is satisfied by the following
form of the action of boosts on single-particle momenta
\begin{equation}
\label{boosts-mom}
\begin{split}
\{N_{i},p_{0}\}= & p_{i} \\
\{N_{i},p_{j}\}= & \delta_{ij}\, p_{0}+\frac{\rho}{2}\left(\epsilon_{ij}\,|\vec{p}|^{2}-\epsilon_{ik}\, p_{k}p_{j}\right) \, .
\end{split}
\end{equation}
Building on these starting points it is easy to complete the description of the relativistic symmetries
of $\rho$-Minkowski by simply insisting that all relevant Jacobi identities are satisfied.
This leads one to the conclusion that the only other needed deformation is in the Poisson bracket between
two boosts,
\begin{equation}
\{N_i,N_j\}=-R\epsilon_{ij}+\frac{\rho}{2}\epsilon_{ij}N_ip_i \, ,
\end{equation}
while all other remaining Poisson brackets are undeformed.

We conclude that the full description of the relativistic symmetries of $\rho$-Minkowski
is given in terms of the following Poisson brackets:
\begin{equation}
\label{symmetry-algebra}
\begin{split}
\{p_{0},p_{i}\}= & 0 \ ,\\
\{p_{i},p_{j}\}= & 0 \ ,\\
\{R,p_{0}\}= & 0 \ ,\\
\{R,p_{i}\}= & \epsilon_{ij}\, p_{j} \ ,\\
\{R,N_{i}\}= & \epsilon_{ij}\, N_{j} \ ,\\
\{N_{i},p_{0}\}= & p_{i} \ ,\\
\{N_{i},p_{j}\}= & \delta_{ij}\, p_{0}+\frac{\rho}{2}\left(\epsilon_{ij}\,|\vec{p}|^{2}-\epsilon_{ik}\, p_{k}p_{j}\right) \ , \\
\{N_{i},N_{j}\}= & -\epsilon_{ij}\, R+\frac{\rho}{2}\epsilon_{ij}\, N_{a}p_{a} \ .
\end{split}
\end{equation}
Notably the ordinary special-relativistic mass Casimir,
\begin{equation}
C=p_0^2-|\vec{p}|^{2} \, ,
\label{goodcasimir}
\end{equation}
is still a Casimir also of the deformed symmetry algebra (\ref{symmetry-algebra}).

\section{Dual-curvature lensing with commuting coordinates}
As for most other studies of the implications of Planck-scale momentum-space curvature,
we shall here focus on the implications of the associated DSR-deformed relativistic symmetries
at the level of the trajectories of particles in the classical limit.
Previous related studies \cite{dsrdesitter}
have shown that, while the noncommutativity of coordinates plays evidently
a crucial role in establishing the form of the DSR-relativistic symmetries, the truly physical content
of the analysis of classical trajectories is the same whether one uses noncommuting coordinates or
commuting ones. In these cases it turns out to be convenient
to perform computations both using commuting and
using noncommuting coordinates, since this facilitates establishing which properties are truly physical
(properties which are coordinate artifacts will change from one formulation to the other).
We therefore study dual-curvature lensing in this section using commuting coordinates, while in the next section
we perform the same study using coordinates with $\rho$-Minkowski Poisson brackets.

The symplectic structure adopted for this section is
\begin{equation}
\label{symplectic-structure}
\begin{split}
\{\mathbf{x}_{i},\mathbf{x_{0}}\}= & 0 \ ,\\
\{\mathbf{x}_{i},\mathbf{x}_{j}\}= & 0 \ ,\\
\{p_{0},\mathbf{x}_{0}\}= & 1 \ ,\\
\{p_{i},\mathbf{x}_{0}\}= & 0 \ ,\\
\{p_{0},\mathbf{x}_{i}\}= & 0 \ ,\\
\{p_{i},\mathbf{x}_{j}\}= & -\delta_{ij} \ .
\end{split}
\end{equation}
It is useful to start by noticing that in terms of this symplectic structure the relativistic-symmetry generators
introduced in the previous section admit the following representation:
\begin{equation}
\label{gen-rep-comm}
\begin{split}
 p_\mu=& p_{\mu} \ ,\\
\mathcal{R}= & \epsilon_{ij}\mathbf{x}_{i}p_{j} \ ,\\
\mathcal{N}_{i}= & \mathbf{x}_{i}p_{0}-\mathbf{x}_{0}p_{i}+\frac{\rho}{2}\epsilon_{ia}\left(\mathbf{x}_{a}|\vec{p}|^{2}-p_{a}\vec{\mathbf{x}}\cdot\vec{p}\right) \ .
\end{split}
\end{equation}
From this representation one finds  that the Poisson brackets between rotations and the spacetime coordinates  are the standard ones, while the Poisson brackets between boosts and spacetime coordinates are deformed:
\begin{equation}
\begin{split}
\!\!\!\!\!\! \{\mathcal{N}_{i},\mathbf{x}_{0}\}= & \mathbf{x}_{i}\{p_{0},\mathbf{x}_{0}\}=\mathbf{x}_{i} \ ,\\
\!\!\!\!\!\! \{\mathcal{N}_{i},\mathbf{x}_{j}\}= & -\mathbf{x}_{0}\{p_{i},\mathbf{x}_{j}\}+\frac{\rho}{2}\epsilon_{ia}\left(\mathbf{x}_{a}\{|\vec{p}|^{2},\mathbf{x}_{j}\}-p_{a}\vec{\mathbf{x}}\cdot\{\vec{p},\mathbf{x}_{j}\}\right)=\\
= & \mathbf{x}_{0}\delta_{ij}+\frac{\rho}{2}\epsilon_{ia}\left(p_{a}\mathbf{x}_{j}-2\mathbf{x}_{a}p_{j}\right) +\frac{\rho}{2}\epsilon_{ij}x_lp_l \ .
\end{split}
\end{equation}
We can of course derive the equations of motion using the mass Casimir (\ref{goodcasimir}) as Hamiltonian
of evolution in an affine parameter, finding
\begin{equation}
\label{eq-motion}
\begin{split}
\dot{\mathbf{x}}_{0}=&\{\mathcal{C},\mathbf{x}_{0}\}=\{p_{0}^{2}-|\vec{p}|^{2},\mathbf{x}_{0}\}=2\, p_{0} \ , \\
\dot{\mathbf{x}}_{i}=&\{\mathcal{C},\mathbf{x}_{i}\}=\{p_{0}^{2}-|\vec{p}|^{2},\mathbf{x}_{i}\}=2\, p_{i} \ , \\
\dot{p}_\mu=&\{\mathcal{C}, p_\mu\}=\{p_{0}^{2}-|\vec{p}|^{2},p_\mu\}=0 \, ,
\end{split}
\end{equation}
where $\dot{q}$ denotes the derivative of $q$ with respect to the affine parameter.
Momenta are constant of motion and
a generic solution of the equations for the worldline can be written as
\begin{equation}
\label{sol-motion}
\mathbf{x}_i-\bar{\mathbf{x}}_i= \frac{p_i}{p_0} (\mathbf{x}_0-\bar{\mathbf{x}}_0) \ ,
\end{equation}
where $\bar{\mathbf{x}}_\mu$ is a point of the worldline.

Our analysis of dual-curvature lensing starts by considering
 an observer Alice that emits from its spacetime origin two massless particles of different energies. The relevant two massless particles
  have spatial momenta $\vec{p}_s=(p_{1s}, 0)$, $\vec{p}_h=(p_{1h}, 0)$, with $p_{1s}>0$, $p_{1h}>0$, and the one with momentum $\vec{p}_h$
  is ``hard" while the one with momentum $\vec{p}_s$
  is ``soft", meaning that $p_{1s} \ll p_{1h}$. From Eq.(\ref{sol-motion}), we have that Alice describes the two particles with the same worldline
\begin{equation}
\label{AliceWorldline}
\begin{split}
\mathbf{x}^A_{1,s}=\mathbf{x}^A_{1,h}= & \mathbf{x}^A_{0} \ ,\\
\mathbf{x}^A_{2,s}=\mathbf{x}^A_{2,h}= & 0 \ .
\end{split}
\end{equation}
Alice is the observer at the source. The main feature of dual-curvature lensing will be manifest in the description give by a distant
detector, with direction-discriminating capabilities, of these
two particles emitted at Alice.
The next step of our analysis is to consider an observer Bob, which is related to Alice by a pure
translation with parameters $b^{\mu}=(b,-b,0)$. Since Bob and Alice are related by a pure translation with parameters $b^{\mu}=(b,-b,0)$,
one has of course the following relationship between Alice's coordinates and Bob's coordinates:
\begin{equation}
\begin{split}
\mathbf{x}_{0}^{B}=&\mathbf{x}_{0}^{A}+b^{\mu}\{p_{\mu},\mathbf{x}_{0}^{A}\}=\mathbf{x}_{0}^{A}+b \ , \\
\mathbf{x}_{1}^{B}=&\mathbf{x}_{1}^{A}+b^{\mu}\{p_{\mu},\mathbf{x}_{1}^{A}\}=\mathbf{x}_{1}^{A}+b \ , \\
\mathbf{x}_{2}^{B}=&\mathbf{x}_{2}^{A}+b^{\mu}\{p_{\mu},\mathbf{x}_{2}^{A}\}=\mathbf{x}_{2}^{A} \ .
\end{split}
\end{equation}
This leads us immediately to Bob's description of the worldlines of the two particles emitted by Alice:
\begin{equation}
\begin{split}
\mathbf{x}^B_{1,s}=\mathbf{x}^B_{1,h}= & \mathbf{x}^B_{0} \, ,\\
\mathbf{x}^B_{2,s}=\mathbf{x}^B_{2,h}= & 0 \, ,
\end{split}
\end{equation}
for which we also used the fact that
 momenta are invariant under pure translations.

Evidently the map between Alice and Bob, purely translated with respect to Alice, remains trivial and unaffected by the curvature
of momentum space. We shall not report explicitly the directional analysis for Bob, but the triviality of the map
clearly implies that there is no dual-curvature lensing in the picture involving these two observers, Alice and Bob.
We shall find that a truly physical feature
(not a coordinate artifact)
of dual-curvature lensing is found in $\rho$-Minkowski when the source and the
 detector are not just purely translated but rather are related by a composition of a translation and a boost.
  It is therefore appropriate to consider at this point a third observer Camilla, purely boosted with respect to Bob. Observer Camilla is
  therefore obtained from Alice by first acting with a translation with parameters $b^\mu=(b, -b, 0)$  and
  then with a boost with parameters $\xi^i=(\xi^{1},0)$
\begin{equation}
\begin{split}
\mathbf{x}^C_\mu =& T_{\xi^i}(T_{b^\mu}(\mathbf{x}^A_\mu))=T_{\xi^i}(x^B_{\mu}) \ , \\
p^C_{\mu} =&T_{\xi^i}(T_{b^\mu}(p^A_\mu))=T_{\xi^i}(p^B_{\mu})\ .
\end{split}
\end{equation}
As a first step toward writing the worldlines of the soft and the hard particle in Camilla's reference frame, we give the transformation laws for momentum between Camilla's frame and Alice's frame, specialized to the case of our interest (in which $p^A_\mu=(p^A_0, p^A_1, 0)$):
\begin{equation}
\begin{split}
\label{Camilla-momenta}
p_{0}^{C}= &p_{0}^{A}+\xi^{1}\{\mathcal{N}_{1},p_{0}^{A}\}=\\
= & p_{0}^{A}+\xi^{1}p_{1}^{A} \ , \\
p_{1}^{C}= &p_{1}^{A}+\xi^{1}\{\mathcal{N}_{1},p_{1}^{A}\}=\\
= & p_{1}^{A}+\xi^{1}p_{0}^{A} \ , \\
p_{2}^{C}= & p_{2}^{A}+\xi^{1}\{\mathcal{N}_{1},p_{2}^{A}\}=\\
= & \xi^{1}\left(\frac{\rho}{2}(p_{1}^{A})^{2}\right) \ .
\end{split}
\end{equation}
Here one sees that the $\rho$-Minkowski deformation of boost transformations is such that, while for Alice the spatial part of the momenta of the soft and hard particles are both directed along the $\mathbf{x}^1$-axis, in Camilla's frame the  momenta have a component also along the $\mathbf{x}^2$-axis, proportional to the energies of the particles in Alice's frame.

Our next step is to determine which coordinates Camilla assigns to the origin of Alice, so
we consider $\bar{\mathbf{x}}^C_\mu=T_{\xi^i}(T_{b^\mu}(\bar{\mathbf{x}}^A_\mu=(0, 0, 0)))$, finding that
\begin{equation}
\begin{split}
\mathbf{x}_{0}^{C}= & b+\xi^{1}b \ , \\
\mathbf{x}_{1}^{C}= & b+\xi^{1}b  \ , \\
\mathbf{x}_{2}^{C}= & 0 \ ,
\end{split}
\end{equation}
which is valid both for the soft particle and for the hard particle. We note that after computing the Poisson brackets we used that $p^A_{2h}=0$ and that $T_{b^\mu 2}=0$.

We are now ready to give the worldlines of the particles according to Camilla. For the soft particle Camilla has
\begin{equation}
\begin{split}
\mathbf{x}_{1,s}^{C}= & \mathbf{x}_{0}^{C} \ ,\\
\mathbf{x}_{2,s}^{C}= & 0 \ ,
\end{split}
\end{equation}
where we neglected the deformation effects (the soft particle is a low-energy particle, and the deformation here of
interest grows with energy), while Camilla's description of the worldline of the hard particle is
\begin{equation}
\begin{split}
\mathbf{x}_{1,h}^{C}= & \mathbf{x}_{0}^{C} \ ,\\
\mathbf{x}_{2,h}^{C}= & \frac{\rho}{2}p_{0,h}^C\xi^{1}\mathbf{x}_{0}^{C} \ .
\end{split}
\end{equation}
This establishes that both the soft particle and the hard particles go through the spacetime origin of Camilla's reference frame,
but we are evidently most interested in what measurement results Camilla would get for the direction of the two particles.
For this purpose it is important to notice that a good measurement procedure for directions involves the setup of a rigid non-point-like
detector: we establish the direction of a particle by seeing that the particle went through two (or more) points of the rigid
extended detector. Since we are thinking of possible applications of our results to the analysis of IceCube-telescope data,
it is nice to notice that IceCube establishes directions exactly in this way.
From a relativist perspective the notion of an extended detector is a bit cumbersome, since it is standard practice to abstract
the notion of a point-like detector in association with the equally abstract notion of relativistic observer.
We can however still rely on these standard abstractions by viewing a rigid extended detector as a network of point-like detectors
in rigid motion. The results will be more transparent by introducing a different relativistic observer for each of the point-like
detectors composing the extended detector, so that each of these many observers has a pointlike detector in the spatial origin of its reference frame.

These conceptual considerations translate in the technical challenge for our computations of introducing two more observers
which we will simply denote with $D$ and $D'$. Both the soft particle and the hard particle go through the origin of Camilla's reference frame,
so if we find that the soft (hard) particle also goes through $D$ ($D'$) then we will have enough information for establishing the
direction of the soft (hard) particle.
It is appropriate to take $D$ as an observer, at rest with respect to Camilla, whose origin of the reference frame coincides with
the point which for Camilla has coordinates $\mathbf{x}^C_{\mu ,D}=(\delta, \delta, 0)$. The observer $D$ is related
to Camilla by a translation with parameters $\delta^{\mu}=(\delta, -\delta, 0)$, so that
\begin{equation}
\begin{split}
\mathbf{x}_{0}^{D}=&\mathbf{x}_{0,D}^{C}+\delta^{\mu}\{p_{\mu},\mathbf{x}_{0,D}^{C}\}=\mathbf{x}_{0,D}^{C}+\delta \ , \\
\mathbf{x}_{1}^{D}=&\mathbf{x}_{1,D}^{C}+\delta^{\mu}\{p_{\mu},\mathbf{x}_{1,D}^{C}\}=\mathbf{x}_{1,D}^{C}+\delta \ , \\
\mathbf{x}_{2}^{D}=&\mathbf{x}_{2,D}^{C}+\delta^{\mu}\{p_{\mu},\mathbf{x}_{2,D}^{C}\}=\mathbf{x}_{2,D}^{C} \ .
\end{split}
\end{equation}
This straightforwardly leads us to the description given by observer $D$ of the worldlines of the soft particle and of the hard particle:
\begin{equation}
\begin{cases}
\mathbf{x}_{1,s}^{D}= & \mathbf{x}_{0}^{D}\\
\mathbf{x}_{2,s}^{D}= & 0
\end{cases}\qquad\begin{cases}
\mathbf{x}_{1,h}^{D}= & \mathbf{x}_{0}^{D}\\
\mathbf{x}_{2,h}^{C^{\prime}}= & \frac{\rho}{2}p_{0,h}^{D}\xi^{1}(\mathbf{x}_{0}^{D}-\delta )
\end{cases}
\end{equation}
where $p_{0,h}^{D}=p_{0,h}^{C}$, since observer Camilla and observer $D$ are connected by a pure translation. Notice that
these equations for the worldlines show in particular that the soft particle goes through the origin of observer $D$ but
the hard particle doesn't.

Next it is convenient to consider another observer $D'$,
also at rest with respect to Camilla, whose origin of the reference frame coincides with
the point which for Camilla has coordinates $\mathbf{x}^C_{\mu ,D'}=(\delta, +\delta, +\frac{\rho}{2}p_{0,h}^{C}\xi^1\delta)$.
The observer $D'$ is related
to Camilla by a translation with parameters $\delta^\mu = (\delta, -\delta, -\frac{\rho}{2}p_{0,h}^{C}\xi^1\delta)$,
so that we have
\begin{equation}
\begin{split}
\mathbf{x}_{0}^{D'}=&\mathbf{x}_{0,D'}^{C}+\delta^{\mu}\{p_{\mu},\mathbf{x}_{0,D'}^{C}\}=\mathbf{x}_{0,D'}^{C}+\delta \ , \\
\mathbf{x}_{1}^{D'}=&\mathbf{x}_{1,D'}^{C}+\delta^{\mu}\{p_{\mu},\mathbf{x}_{1,D'}^{C}\}=\mathbf{x}_{1,D'}^{C}+\delta \ , \\
\mathbf{x}_{2}^{D'}=&\mathbf{x}_{2,D'}^{C}+\delta^{\mu}\{p_{\mu},\mathbf{x}_{2,D'}^{C}\}=\mathbf{x}_{2,D'}^{C} +\frac{\rho}{2}p_{0,h}^{C}\xi^1\delta\ .
\end{split}
\end{equation}
Therefore according to observer $D'$ the worldlines of the soft particle and of the hard particle are described by
\begin{equation}
\begin{cases}
\mathbf{x}_{1,s}^{D'}= & \mathbf{x}_{0}^{D'} \ ,\\
\mathbf{x}_{2,s}^{D'}= & \frac{\rho}{2}p_{0,h}^{D'}\xi^{1}\delta \ ,
\end{cases}\qquad
\begin{cases}
\mathbf{x}_{1,h}^{C''}= & \mathbf{x}_{0}^{D'} \ ,\\
\mathbf{x}_{2,h}^{C''}= & \frac{\rho}{2}p_{0,h}^{D'}\xi^{1}\mathbf{x}_{0}^{D'} \ .
\end{cases}
\end{equation}
where $p_{0,h}^{D'}=p_{0,h}^{D}=p_{0,h}^{C}$, since the observers Camilla, $D$ and $D'$
 are all connected by a pure translation. Notice that
these equations for the worldlines show in particular that the hard particle goes through the origin of observer $D$ but
the soft particle doesn't.

So we found that the soft particle goes through $D$ and Camilla, while the hard particle goes through
$D'$ and Camilla. This is the dual-curvature lensing that we were looking for: the particle are emitted by the
same source but their detection manifests a different direction of propagation.
From the fact that the soft particle goes through $D$ and Camilla and the hard particle goes through
$D'$ and Camilla one easily finds that the angle $\theta$ characterizing the difference in their observed
directions of propagation is given by
\begin{equation}
tan\theta \simeq
\frac{\rho}{2}p_{0,h}^{C}\xi^1 \ .
\end{equation}

\section{Dual-curvature lensing with $\rho$-Minkowski commuting coordinates}
As announced earlier, we shall redo, now assuming $\rho$-Minkowski
Poisson brackets, the analysis done in the previous section with ``commutative" Poisson brackets.
We shall find the same physical result for dual-curvature lensing, but several details of the
derivation will be different.

So we now adopt the following symplectic structure:
\begin{equation}
\label{newsymplectic}
\begin{split}
\{x_{i},x_{0}\}= & \rho\epsilon_{ij}\, x_{j} \ , \\
\{x_{i},x_{j}\}= & 0 \ , \\
\{p_{0},x_{0}\}= & 1 \ ,\\
\{p_{0},x_{i}\}= & 0 \ ,\\
\{p_{i},x_{0}\}= & 0 \ ,\\
\{p_{i},x_{j}\}= & -\delta_{ij}+\rho\epsilon_{ij}\, p_{0} \ .
\end{split}
\end{equation}
The fact that the physical results cannot depend on the choice between this symplectic structure and the symplectic structure
used in the previous section is clear upon noticing that these two symplectic structures are connected
by a simple momentum-dependent redefinition of spacetime coordinates:
\begin{equation}
\label{map}
\begin{split}
\mathbf{x}_{0}= & x_{0} \ ,\\
\mathbf{x_{i}=} & x_{i}-\rho\epsilon_{ij}\, x_{j}p_{0} \ .
\end{split}
\end{equation}
where $\mathbf{x}_{\mu}$ are the ``commutative"  coordinates used in the previous section, while the $x_\mu$ are used
in this section.

In terms of the  $x_\mu$ coordinates
the representation of the generators of the relativistic symmetries is
\begin{equation}
\begin{split}
 p_\mu =& p_{\mu} \ ,\\
\mathcal{R}= & \epsilon_{ij}x_{i}p_{j}-\rho\, p_{0}\vec{x}\cdot\vec{p} \ ,\\
\mathcal{N}_{i}= & x_{i}p_{0}-x_{0}p_{i}+\frac{\rho}{2}\epsilon_{ij}\left(x_{j}|\vec{p}|^{2}-p_{j}\vec{x}\cdot\vec{p}-2\, x_{j}p_{0}^2\right) \ .
\end{split}
\end{equation}
Accordingly we then have that
\begin{equation}
\begin{split}
\{\mathcal{R},x_{0}\}= & \{\epsilon_{ij}x_{i}p_{j}-\rho\, p_{0}x_{i}p_{i},x_{0}\}=0 \ ,\\
\{\mathcal{R},x_{i}\}= & \{\epsilon_{ab}x_{a}p_{b}-\rho\, p_{0}x_{a}p_{a},x_{i}\}=\\
= & \epsilon_{ab}x_{a}\{p_{b},x_{i}\}-\rho\, p_{0}x_{a}\{p_{a},x_{i}\}=\\
= & \epsilon_{ab}x_{a}(-\delta_{bi}+\rho\epsilon_{bi}\, p_{0})+\rho\, p_{0}x_{a}\delta_{ai}=\\
= & \epsilon_{ia}x_{a}+\rho\,(-x_{i}p_{0}+p_{0}x_{i})=\\
= & \epsilon_{ia}x_{a} \ ,
\end{split}
\end{equation}
\begin{equation}
\begin{split}
\{\mathcal{N}_{i},x_{0}\}= & \{x_{i}p_{0}-x_{0}p_{i}+\frac{\rho}{2}\epsilon_{ia}\left(x_{a}|\vec{p}|^{2}-p_{a}\vec{x}\cdot\vec{p}-2\, x_{a}p^2_{0}\right),x_{0}\}=\\
= & x_{i}\{p_{0},x_{0}\}+p_{0}\{x_{i},x_{0}\}-\frac{\rho}{2}\epsilon_{ia}2\, x_{a}\{p^2_{0},x_{0}\}=\\
= & x_{i}-\rho\epsilon_{ij}\, x_{j}p_0 \ ,\\
\{\mathcal{N}_{i},x_{j}\}= & \{x_{i}p_{0}-x_{0}p_{i}+\frac{\rho}{2}\epsilon_{ia}\left(x_{a}|\vec{p}|^{2}-p_{a}\vec{x}\cdot\vec{p}-2\, x_{a}p^2_{0}\right),x_{j}\}=\\
& \!\!\!\!\!\!\!\!\!\!\!\!\!\!\!\!\!\!\!\!\!\!\!\! =  -x_{0}\{p_{i},x_{j}\} -p_{i}\{x_{0},x_{j}\}+\frac{\rho}{2}\epsilon_{ia}\left(x_{a}\{|\vec{p}|^{2},x_{j}\}-x_{b}\{p_{a}p_{b},x_{j}\}\right)=\\
& \!\!\!\!\!\!\!\!\!\!\!\!\!\!\!\!\!\!\!\!\!\!\!\!\!\!\!\!\!\!\!\!\!\!\!\!\!\!\!\!\!\!\!\! =  -x_{0}(-\delta_{ij}+\rho\epsilon_{ij}\, p_{0})+\rho\epsilon_{jk}x_k p_i+\frac{\rho}{2}\epsilon_{ia}\left(-2x_{a}p_{j}+x_{b}(\delta_{bj}p_{a}+\delta_{aj}p_{b})\right)=\\
& \!\!\!\!\!\!\!\!\!\!\!\!\!\!\!\!\!\!\!\!\!\!\!\!\!\!\!\!\!\!\!\!\!\!\!\! =  \delta_{ij}x_{0}+\rho\,\left(\frac{1}{2}(\epsilon_{ia}p_{a}x_{j}+\epsilon_{ij}x_{a}p_{a})+\epsilon_{jk}x_k p_i-\epsilon_{ia}x_{a}p_{j}-\epsilon_{ij}p_{0}x_{0}\right) \ .
\end{split}
\end{equation}
It is equally straightforward to see that the equations of motion take the form
\begin{equation}
\begin{split}
\dot{x}_{0}= &\{\mathcal{C},x_{0}\}=\{p_{0}^{2}-|\vec{p}|^{2},x_{0}\}=2p_{0} \ , \\
\dot{x}_{i}=&\{\mathcal{C},x_{i}\}=\{p_{0}^{2}-|\vec{p}|^{2},x_{i}\}=-2p_{a}\{p_{a},x_{i}\}=2p_{i}+2\rho\epsilon_{ij}p_{j}p_{0} \ , \\
\dot{p}_\mu=&\{\mathcal{C},x_{0}\}=\{p_{0}^{2}-|\vec{p}|^{2},x_{0}\}=0 \ .
\end{split}
\end{equation}
This implies that  momenta are constant of motion and that a generic  worldline can be written as
\begin{equation}
\label{worldlinecompact}
x_i-\bar{x}_i=\left(\frac{p_{i}}{p_{0}}+\rho\epsilon_{ij}p_{j}\right)(x_0-\bar{x}_0) \ ,
\end{equation}
where $p_\mu$ is the momentum of the particle and $\bar{\mathbf{x}}_\mu$ is a point of the worldline.

Of course we shall focus again on the same two massless particles emitted at Alice already considered in the previous section.
With the choice of coordinates made in this section one has that Alice's description of the relevant two worldlines
is
\begin{equation}
\label{gac99}
\begin{cases}
x_{1,s}^{A}=&x_{0}^{A} \ ,\\
x_{2,s}^{A}=&0 \, ,
\end{cases}\qquad \begin{cases}
x_{1,h}^{A}=&x_{0}^{A} \ ,\\
x_{2,h}^{A}=&-\rho\, p_{0}x_{0}^{A} \, ,
\end{cases}
\end{equation}
where $p_{0}$ denotes the energy of the hard particle ($p_{0}\equiv p_{0,h}$)
while in the equation for $x_{2,s}^{A}$ we neglected a term going with the energy
of the soft particle (we are assuming again that for the soft particle the $\rho$-dependent
effects will be negligible).
Notice that, even though the two particles have $p_{2,s}=p_{2,s}=0$, in the coordinatization used in
this section one gets that $x_2$ changes over time (visible in (\ref{gac99}) for the hard particle, neglected for the soft particle).
This is just a coordinate artifact due to the symplectic structure
(\ref{newsymplectic}), and indeed even in this coordinatization the particles reach Bob (the observer already introduced in the
previous section, which is purely translated with respect to Alice along the $x_1$ direction).
In order to verify this important fact we start by characterizing the transformation between Alice's coordinates and
Bob coordinates, using the coordinatization of this section, taking into account that again the translation parameters
connecting Alice and Bob are $b^{\mu}=(b,-b,0)$:
\begin{equation}
\begin{split}
x_{0}^{B}=&x_{0}^{A}+b^{\mu}\{p_{\mu},x_{0}^{A}\}=x_{0}^{A}+b \ , \\
x_{1}^{B}=&x_{1}^{A}+b^{\mu}\{p_{\mu},x_{1}^{A}\}=x_{1}^{A}+b \ , \\
x_{2}^{B}=&x_{2}^{A}+b^{\mu}\{p_{\mu},x_{2}^{A}\}=x_{2}^{A}-\rho b p^A_{0} \ .
\end{split}
\end{equation}
Equipped with this one easily arrives at  Bob's of the worldlines of the soft particle and of the hard particle:
\begin{equation}
\begin{split}
x_{1,s}^{B}= & x_{0}^{B} \ ,\\
x_{2,s}^{B}= & 0 \ .
\end{split}
\end{equation}
\begin{equation}
\begin{split}
x_{1,h}^{B}= & x_{0}^{B} \ ,\\
x_{2,h}^{B}= & -\rho p^{B}_{0,h}x_{0}^{B} \ .
\end{split}
\end{equation}
This shows that indeed both the soft particle and the hard particle go through Bob's spacetime origin. With the ``$\rho$-Minkowski coordinates" we are using in this section there is a coordinate artifact apparently attributing to the particles a motion also along the $x_2$ direction, but physically our two  particles just move along the 1 direction and indeed they are emitted at Alice and they are detected at Bob, with Bob
purely translated with respect to Alice along the 1 direction.

The next step toward our goal of characterizing dual-curvature lensing with $\rho$-Minkowski coordinates is again the introduction
of observer Camilla, the observer  Camilla which is purely boosted  with respect to Bob (and therefore obtained from Alice by performing first a translation and then a boost):
\begin{equation}
\begin{split}
x^C_\mu =& T_{\xi^i}(T_{b^\mu}(x^A_\mu))= T_{\xi^i}(x_{\mu}^B) \ , \\
p^C_{\mu} =&T_{\xi^i}(T_{b^\mu}(p^A_\mu))=T_{\xi^i}(p^B_{\mu})\ ,
\end{split}
\end{equation}
where $b^\mu=(b, -b, 0)$ and $\xi^i=(\xi^{1},0)$.

The trasformation laws for momenta between Alice's frame and Camilla's frame,
specialized to the case of our interest, in which the spatial momentum of the particles does not have a component along the $x_2$ direction,
 take the following form:
\begin{equation}
\begin{split}
\nonumber
p_{0}^{C}= &p_{0}^{A}+\xi^{1}\{\mathcal{N}_{1},p_{0}^{A}\}=\\
= & p_{0}^{A}+\xi^{1}p_{1}^{A} \ , \\
p_{1}^{C}= &p_{1}^{A}+\xi^{1}\{\mathcal{N}_{1},p_{1}^{A}\}=\\
= & p_{1}^{A}+\xi^{1}p_{0}^{A} \ , \\
p_{2}^{C}= & p_{2}^{A}+\xi^{1}\{\mathcal{N}_{1},p_{2}^{A}\}=\\
= & \xi^{1}\left(\frac{\rho}{2}(p_{1}^{A})^{2}\right) \ .
\end{split}
\end{equation}
Then proceeding just with the same steps of derivation discussed in the previous section on easily arrives
at the description of the worldlines accoring to Camilla:
\begin{equation}
\begin{cases}
x_{1,s}^{C}=&x_{0}^{C} \ ,\\
x_{2,s}^{C}=&0 \ ,
\end{cases}\qquad \text{hard}\begin{cases}
x_{1,h}^{C}=&x_{0}^{C} \ ,\\
x_{2,h}^{C}=&-\rho\, p^{C}_{0,h}x_{0}^{C} +\frac{\xi^1}{2}\rho p_{0,h}^C x_0^C\ .
\end{cases}
\end{equation}
Importantly both the soft particle and the hard particle go through the spacetime origin of Camilla's reference frame.

Toward our goal of describing dual-curvature lensing with $\rho$-Minkowski coordinates,
we are left with only the final step, the one involving the observers $D$ and $D'$.
The steps of derivation are the same as in the previous section. Some formulas take a different form but the conclusions
for the physical feature of dual-curvature lensing is unchanged.
One easily finds that according to observer $D$ the worldlines take the form
\begin{equation}
\begin{cases}
x_{1,s}^{D}=&x_{0}^{D} \ ,\\
x_{2,s}^{D}=&0 \ ,
\end{cases}\qquad \begin{cases}
x_{1,h}^{D}=&x_{0}^{D} \ ,\\
x_{2,h}^{D}=&-\rho\, p^{D}_{0,h}x_{0}^{D} +\frac{\rho}{2}\xi^1 p_{0,h}^{D} x_0^{D}-\frac{\rho}{2}\xi^1 \delta p_{0,h}^{D}\, ,
\end{cases}
\end{equation}
which again show that  the soft particle goes through the spacetime origin of observer $D$, while the hard particle does not.

Observer $D'$ describes the same worldlines as follows
\begin{equation}
\begin{cases}
x_{1,s}^{D'}=&x_{0}^{D'} \ ,\\
x_{2,s}^{D'}=& \frac{\rho}{2}p_{0,h}^{D'}\xi^1\delta \, ,
\end{cases}\qquad \text{hard}\begin{cases}
x_{1,h}^{D'}=&x_{0}^{D'} \ ,\\
x_{2,h}^{D'}=&-\rho\, p^{D'}_{0,h}x_{0}^{D'} +\frac{\rho}{2}\xi^1 p_{0,h}^{D'} x_0^{D'}\ .
\end{cases}
\end{equation}
which again show that the hard particle goes through the spacetime origin of observer $D$, while the soft particle does not.

Finally reasoning just as done at the end of the previous section one finds from these results
that the angle $\theta$ characterizing the difference in their observed
directions of propagation is given by
\begin{equation}
tan\theta \simeq \frac{\rho}{2}p_{0,h}^{C}\xi^1 \ .
\end{equation}
This is the same result found at the end of the previous section. While some intermediate steps of derivation do depend on
the choice of coordinatization, the result for dual-curvature lensing is the same independently on whether one uses ``commuting coordinates",
as done in the previous section, or ``$\rho$-Minkowski coordinates", as done in this section.

\section{Closing remarks}
We here made some significant steps toward the understanding of dual-curvature lensing,
most notably establishing as a truly physical effect.
The noncommutative $\rho$-Minkowski turned out to be ideally suited for
investigating dual-curvature lensing at the classical-mechanics level,
and it should naturally provide a good option for exploring
quantum manifestations of dual-curvature lensing,
in the sense of studies of dual-curvature redshift such as the one reported in Ref.\cite{fuzzy1}.
Of course, ultimately the most significant applications will concern phenomenology.
In this respect it is noteworthy that
for $\rho$-Minkowski we found dual-curvature lensing amounting to a mismatch of
directions given by an angle $\theta$ of order $\rho p_0 \xi$,
with $\rho$ the noncommutativity scale, $p_0$ the energy difference between the two particles,
and $\xi$ the boost between the (distant) observers at the source and at the detector.
We expect that the quantification of
dual-curvature lensing might be different in different spacetimes, so, in order to
give guidance to the relevant phenomenology, it would be important to find
other quantum spacetimes, in addition to $\rho$-Minkowski,
to be used in the exploration of the realm of possibilities for
 dual-curvature lensing.

\end{document}